\documentclass[
aps,%
12pt,%
final,%
notitlepage,%
oneside,%
onecolumn,%
nobibnotes,%
nofootinbib,
superscriptaddress,%
noshowpacs,%
centertags]%
{revtex4}

\usepackage{graphicx}
\usepackage[english]{babel}
\begin{document}
%
\title{Two-Particle Correlations in the Wave Function and
       Covariant Current Approaches}

\author{\firstname{Dmitry} \surname{Anchishkin}}
\email[anch@bitp.kiev.ua]{}
\affiliation{ Bogolyubov Institute for Theoretical Physics, 03068
              Kiev, Ukraine }

\author{\firstname{Ulrich} \surname{Heinz}}
\email[heinz@mps.ohio-state.edu]{}
\affiliation{ Physics Department, The Ohio State University,
              Columbus, OH 43210 }
%

\begin{abstract}
We consider two-particle correlations, which appear in relativistic
nuclear collisions due to the quantum statistics of identical
particles, in the frame of two formalisms: wave-function and
current.
The first one is based on solution of the Cauchy problem,
whereas the second one is a so-called current parametrization of the
source of secondary particles.
We argue that these two parameterizations of the source coincide when
the wave function at freeze-out times is put in a specific
correspondence with a current.
Then, the single-particle Wigner density evaluated in both
approaches gives the same result.

\end{abstract}

\date{\today}

\maketitle
%
\section{Introduction}
\label{sec1}

The models and approaches which are used to
describe the processes occurred in the reaction region in relativistic
heavy-ion collisions are examined
by comparison of provided predictions with experimental data on
single-, two- and many-particle momentum spectra, which contain
information about the source at the early stage (photons, dileptons)
and at the stage of so called ``{\it freeze-out}'' (hadron spectra).
Two-particle correlations or the Hanbury-Brown-Twiss interferometry
(HBT) encapsulate information about the space-time
structure and dynamics of the emitting source
\cite{GKW,boal,heinz99,weiner-2000,padula-2004,pratt-2005,csorgo-2005}.
Usually, consideration of the correlations, which occur in relativistic
heavy-ion collisions, assumes that:
(i)  the particles are emitted independently (or the source is
     completely chaotic), and
(ii) finite multiplicity corrections can be neglected.
Both approximations are expected to be good for high energy nuclear
collisions with large multiplicities.
Then, correlations reflect
a) the effects from symmetrization (antisymmetrization) of the
amplitude to detect identical particles with certain momenta, and
b) the effects which are generated by the final state interactions of the
detected particles between themselves and with the source.
On the first sight one can regard the final state interactions (FSI)
as a contamination of ``pure'' particle correlations.
But, it should be noted that the FSI depend on
the structure of the emitting source and thus provide as well
information about source dynamics   \cite{anch98}.

Several surprising questions motivated by new experimental data
appeared recently in the HBT.
For instance, the experimental measurements on two-pion correlations
\cite{STAR-2001,PHENIX-2002,PHENIX-2004,STAR-2005}
give the ratio of $R_{\rm out}/R_{\rm side} \approx 1$, what is much
smaller than
that predicted theoretically (the so called ``RHIC HBT Puzzle'').
This raises the question to what extent some of the model predictions
are consistent with experimental measurements
\cite{heinz-2006,stocker-2006} or may be the observed discrepancies
are due to such an ``apples-with-oranges'' comparison.
All this drew attention and inspired a more detailed discussion of
the theoretical background of the HBT.
In the present paper we are going along this line, we would like to
clarify a question concerning
different kinds of  parametrization exploited in the HBT.

The nominal quantity expressing the correlation function in terms
of experimental distributions \cite{boal} is
\begin{equation}
C({\bf k}_a,{\bf k}_b)=
\frac{\displaystyle P_2\left({\bf k}_a, {\bf k}_b\right) }
{\displaystyle P_1\left({\bf k}_a\right) \,
P_1\left({\bf k}_b\right) }
\ ,
\label{i1}
\end{equation}
\noindent where
$P_1\left({\bf k}\right) =E\, d^3N /d^3k$
and
$P_2\left({\bf k}_a, {\bf k}_b\right) = E_a \, E_b \ d^6N /(d^3k_ad^3k_b)$
are single- and two-particle cross-sections.

In the absence of the final state interactions the theoretical expression
for the two-particle correlator reads
\begin{equation}
C(q,K) = 1\,   \pm  \,
\frac{\displaystyle
\left| \int  d^4 X \, e^{i  q\cdot X } S(X,K) \right| ^2 }
{\displaystyle
 \int  d^4 X \, S\left( X,K+\frac{q}{2} \right) \,
 \int  d^4 Y \, S\left( Y,K-\frac{q}{2} \right)
}
\ ,
\label{i2}
\end{equation}
where  $K=(k_a+k_b)/2 \ , \ \ q=k_a-k_b$.
This expression was obtained in the different approaches.
In the so called "wave-function" approach \cite{anch98}
source function $S(X,K)$ is defined in the following way
 \begin{equation}
  S_{\rm wf}(X,K) = \int d^4x\, e^{i K\cdot x}\,
  \sum_{\gamma , \gamma '} \rho_{\gamma \gamma '}\,
  \psi_\gamma \left(X+{\textstyle{x\over 2}}\right) \,
  \psi_{\gamma '}^*\left(X-{\textstyle{x\over 2}}\right)
\, ,
 \label{i3}
 \end{equation}
where $\rho_{\gamma \gamma '}$ is the density matrix which in
thermal equilibrium has the form
$\rho_{\gamma \gamma '}=\delta_{\gamma \gamma '} \exp{(-E_\gamma/T)}$.
The wave function, $\psi_\gamma \left( t,{\bf x} \right)$, is  taken
at freeze-out times, i.e. $t \in t_\Sigma$.
Freeze-out hyper-surface $\Sigma$ is a spatial surface which moves in
space in the same way as, for instance, the surface of the balloon
during pumping.
It represents an imaginary border between two domains:
inside the surface a strong dynamics takes place whereas outside the
surface the particles propagate outward freely.
Wave function at freeze-out times can be regarded as initial one for
its further history and because its further evolution is free (we do
not discuss final state interactions so far) it can be easily taken
into account.
As it intuitively understood the free evolution can be reverse back
and resulting cross-section and other measurable physical quantities,
for instance source function $S(X,K)$,
are determined through initial values of the wave function, i.e.
by the values of the wave function at freeze-out times.
Rigorous evaluations give exactly this result.
On the other hand, the strong dynamics which acts inside freeze-out
hyper-surface results in creation of the quantum state at freeze-out
times.
Hence, the wave function at freeze-out times is a final state of the
strong dynamics.
Representing experimentally measured quantities with the help of these
states we can study strong interactions in dense and hot nuclear
matter.
Because of this creativity the separation of the interaction
scales in space and time which is made with the help of freeze-out
hyper-surface looks so attractive.

Correlation function (\ref{i2}) was derived first in the model where
essential point is a parametrization of the source by use of the
currents $J_\gamma (x)$ \cite{GKW} (see also \cite{chapman94}) which
become then the constituent elements of the source function
 \begin{equation}
  S_{\rm cur}(X,K) = \int d^4x\, e^{i K\cdot x}\,
  \sum_{\gamma , \gamma '} \rho_{\gamma \gamma '}\,
  J_\gamma \left(X+{\textstyle{x\over 2}}\right) \,
  J_{\gamma '}^* \left(X-{\textstyle{x\over 2}}\right) \, .
 \label{i4}
 \end{equation}

As a matter of fact, both approaches should give the same result in
the region where they are valid.
The goal of this paper is to find relation between source functions
(\ref{i3}) and (\ref{i4}) obtained in wave function approach and
covariant current approach respectively.

\section{Single- and two-particle cross sections without FSI}
\label{sec2}
In this section we consider the two-particle quantum statistical
correlations when one neglects the final state interactions of the
detected particles.
This phenomenon is visualized most transparently on the bases of the
standard quantum mechanics.
First, we briefly consider the so called wave function parametrization of
the source in nonrelativistic approach.
This approach allows one to include also into consideration the final
state interactions \cite{anch98}.
Relativistic picture is considered on the base of the current
parametrization and then on the base of the wave function
parametrization of the source.
First, we compare these two approaches in a non-relativistic sector
and put in correspondence the source functions (\ref{i3}) and (\ref{i4}).
After that the same comparison is carried out for relativistic sector.

\subsection{ Wave function parametrization of the source.
             Nonrelativistic approach }
\label{sec2-1}

The probability to register two-particles which are created in the
relativistic heavy ion collisions and have definite asymptotic momenta
${\bf k}_a$ and ${\bf k}_b$
is compared usually with the probability to register independently two
particles with the same momenta.
That is why, we first turn to consideration of the single-particle
spectrum.

\begin{figure}
{\includegraphics[width=11cm, height=8cm, angle=0]{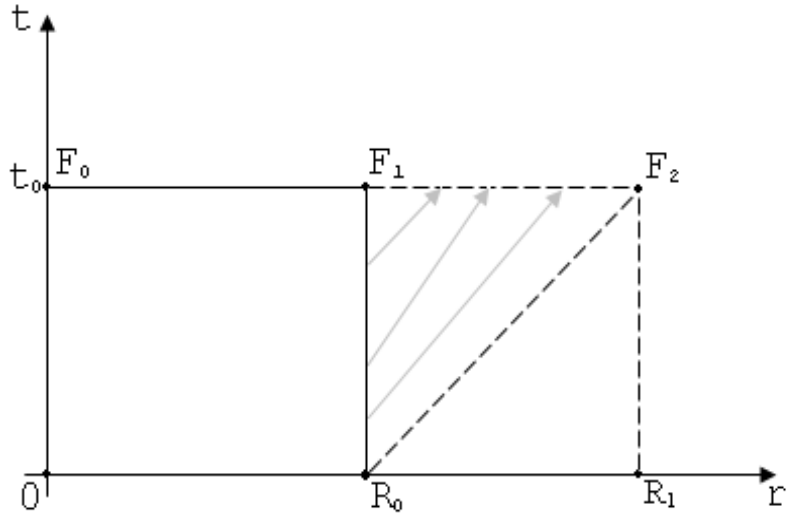} }
\caption{
Sudden freeze-out hyper-surface $F_0 F_1 R_0$ for spherically
symmetric fireball.}
\label{figure:const-t}
\end{figure}

Let us consider a single-particle state $\psi_\gamma$ emitted by the
source.
Its propagation to the detector is governed by the Schr\"odinger
equation
 \begin{equation}
  i \frac{\partial \psi_\gamma({\bf x},t)}
         {\partial t} =
  \hat{h}({\bf x})\,
  \psi_\gamma({\bf x},t) \, ,
 \label{wf-1}
 \end{equation}
where
$  \hat{h} ({\bf x}) = - \frac{1}{2m} {\bf \nabla}^2 $.
The index $\gamma$ denotes a complete set of 1-particle quantum numbers.
Equation~(\ref{wf-1}) is solved by
$
  \psi_\gamma({\bf x},t,t_0) =
  \exp{ [ - i \hat{h}({\bf x}) (t-t_0) ]}\,
  \psi_\gamma({\bf x},t_0)
$
in terms of the single-particle wave function at some initial time
$t_0$, see Fig.\ref{figure:const-t}.
For the spherically symmetric fireball the values of the wave function
$\psi_\gamma({\bf x},t_0)$ parameterize the
``freeze-out distribution'' of the particles inside the sphere of
the radius $R_1$ as it is depicted in Fig.\ref{figure:const-t}.
%
%
We assume that the detector measures asymptotic momentum eigenstates, i.e.
that it acts by projecting the emitted single-particle state onto
$
  \phi^{\rm out}_{\bf k}({\bf x},t)
  = \exp{[i {\bf k}\cdot {\bf x} - i\omega ({\bf k}) t]}
$ ,
where $\omega ({\bf k}) =  {\bf k}^2/2m$.
The measured single-particle momentum amplitude is then
 \begin{equation}
   A_\gamma({\bf k},t_0) = \lim_{t\to\infty}
   \int d^3x\,
   \phi^{\rm out,*}_{\bf k} ({\bf x},t) \,
   \psi_\gamma({\bf x},t,t_0) \, .
 \label{wf-2}
 \end{equation}
The single-particle probability to detect the particle with certain momentum
is obtained by averaging (\ref{wf-2})
and its complex conjugate with the density matrix
$\rho _{\gamma \gamma'}$ defining the source dynamics.
This density matrix is characterized by a probability distribution for
the single-particle quantum numbers $\gamma $ and by a distribution of
emission times $t_0$. We write
 \begin{equation}
   P_1({\bf k}) =
   \sum _{\gamma ,\,\gamma '}
    \rho _{\gamma \gamma '}\,
   A_{\gamma '}  \left( {\bf k},t_0 \right)\,
   A^*_{\gamma } \left( {\bf k},t_0 \right)
   \, .
 \label{wf-3}
 \end{equation}
%
We define the single particle Wigner density of the source
which with accounting for emission times, $t=t_0$, reads
\begin{equation}
S\left(\frac{x_1+x_2}{2},K\right)
=
\int d^4(x_1-x_2)\, \delta(x_1^0-t_0) \, \delta(x_2^0-t_0) \,
  e^{i K\cdot (x_1-x_2)}\,
  \sum_{\gamma ,\, \gamma '} \rho_{\gamma \gamma '}\,
  \psi_\gamma (x_1) \,  \psi_{\gamma '}^*(x_2) \, .
\label{i3a}
\end{equation}
This function accumulates all information about the source which emits
the particles.
Making transformation to new coordinates,
$X=(x_1+x_2)/2, \ \ x=x_1-x_2$, the source function gets the form
\begin{equation}
S(X,K)
=
\delta(X_0-t_0)\int d^4x\, \delta(x_0) \, e^{i K\cdot x}\,
  \sum_{\gamma ,\, \gamma '} \rho_{\gamma \gamma '}\,
  \psi_\gamma \left(X+{\textstyle{x\over 2}}\right) \,
  \psi_{\gamma '}^*\left(X-{\textstyle{x\over 2}}\right)
\, .
\label{i3b}
\end{equation}
Then, the expression for the single-particle spectrum (\ref{wf-3}) can
be rewritten with making use of the source function
 \begin{equation}
   P_1({\bf K}) = \int d^4X\, S(X,K)
  \, .
 \label{wf-4}
 \end{equation}
Note, the factor $\delta(X_0-t_0) \delta(x_0)$ in (\ref{i3b})
carries information about space-like hypersurface where initial values
of the wave function are given.
For the sake of simplicity of the general scheme we start our
consideration from a flat hypersurface, $t=t_0$, depicted in
Fig.\ref{figure:const-t}.
We turn to an arbitrary hypersurface in the next sections where
relativistic approach is elaborated.

\medskip

Let us consider a two-particle state $\psi_\gamma$ emitted by the
source.
Its propagation to the detector is governed by the Schr\"odinger
equation
 \begin{equation}
  i \frac{\partial \psi_\gamma({\bf x}_a,{\bf x}_b,t)}
         {\partial t} =
  \hat{H}({\bf x}_a,{\bf x}_b)\,
  \psi_\gamma({\bf x}_a,{\bf x}_b,t) \, ,
 \label{wf-5}
 \end{equation}
where
$\hat{H}({\bf x}_a,{\bf x}_b) = \hat{h} ({\bf x}_a) + \hat{h} ({\bf x}_b)$.
The index $\gamma$ denotes a complete set of 2-particle quantum numbers.
Equation~(\ref{wf-5}) is solved by
 \begin{equation}
  \psi_\gamma({\bf x}_a,{\bf x}_b,t) =
  \exp{[- i \hat{H}({\bf x}_a,{\bf x}_b) (t-t_0)]}\,
  \psi_\gamma({\bf x}_a,{\bf x}_b,t_0) \, ,
 \label{wf-5a}
 \end{equation}
in terms of the two-particle wave function at some initial time $t_0$.
Detector acts by projecting the emitted two-particle state onto
$
  \phi^{\rm out}_{{\bf p}_a,{\bf p}_b}({\bf x}_a,{\bf x}_b,t)
  = e^{i ({\bf p}_a\cdot {\bf x}_a - \omega_a t)}\,
    e^{i ({\bf p}_b\cdot {\bf x}_b - \omega_b t)}
$ ,
where
$\omega_{a,b} = {\bf p}_{a,b}^2/2m^2$.
We will only consider the case of pairs of identical particles, $m_a=m_b=m$.
The measured two-particle momentum amplitude is then
 \begin{equation}
   A_\gamma({\bf k}_a,{\bf k}_b) =
   \lim_{t\to\infty}
   \int d^3x_a\, d^3x_b\,
   \phi^{\rm out,*}_{{\bf k}_a,{\bf k}_b} ({\bf x}_a,{\bf x}_b,t) \,
   \psi_\gamma({\bf x}_a,{\bf x}_b,t) \, .
 \label{wf-6}
 \end{equation}

We assume the two particles are emitted independently, implying that at
some freeze-out time $t_a$ the two-particle wave function
$\psi_\gamma({\bf x}_{a},{\bf x}_{b},t)$ factorizes
 \begin{equation}
   \psi_\gamma({\bf x}_a,{\bf x}_b,t_a) = \frac{1}{\sqrt{2}}
   \left[ \psi_{\gamma_a}({\bf x}_a,t_a)\,
          \psi_{\gamma_b}({\bf x}_b,t_a) \,
   \pm \,
          \psi_{\gamma_a}({\bf x}_b,t_a)\,
          \psi_{\gamma_b} ({\bf x}_a,t_a) \right]
\, .
 \label{wf-7}
 \end{equation}
The indices $\gamma_a,\gamma_b$ on the single-particle wave functions
now label complete sets of single-particle quantum numbers.
The time moment $t_a=t_0$ is the emission time of the latest emitted
particle.
Because of the symmetry of the wave function (\ref{wf-7}) it does not
matter what time is nominated as latest one, $t_a$ or $t_b$.
By this we assume that symmetrization occurs when the last of the two
particles is frozen out from a strongly interacting bulk.

After hermitian inversion of the evolution operator and applying it to
symmetrised (antisymmetrised) out-state two-particle amplitude
(\ref{wf-6}) gets the form
\begin{equation}
   A_{\gamma_a, \gamma_b}({\bf k}_a,{\bf k}_b,t_0) =
  \frac{1}{\sqrt{2} } \int d^3x_a\, d^3x_b \,
\left[
      e^{-i(k_a\cdot x_a + k_b\cdot x_b) }
       \pm \,
      e^{ -i(k_a\cdot x_b + k_b\cdot x_a) }
\right]^*
\,
      \psi_{\gamma_a}(x_a)\, \psi_{\gamma_b}(x_b)
\ ,
\label{wf-8}
\end{equation}
where $x_a^0=t_0\ \  {\rm and}\ \ x_b^0=t_0$ and by relabeling the
variables of integration we transferred symmetrization from the state
(\ref{wf-7}) onto out-state.
By this we represent the measured two-particle momentum amplitude
as projection of non-symmetrized two-particle wave function taking at
emission times onto symmetrised (antisymmetrised) plane waves taking
as well at emission times.

The two-particle probability to detect two particles with momenta
${\bf k}_a$ and ${\bf k}_b$
is obtained by averaging two-particle amplitude (\ref{wf-8})
and its complex conjugate with the density matrix defining the source.
This density matrix is characterized by a probability distribution for
the two-particle quantum numbers $(\gamma_a,\gamma_b)$,  also we average
by a distribution of emission times $(t_a,\, t_b)$. We write
 \begin{equation}
    P_2({\bf k}_a,{\bf k}_b) =
    \sum  _{\gamma_a \gamma_b,\gamma_{a'} \gamma_{b'} }
   \rho _{\gamma_a \gamma_{a'}}\, \rho _{\gamma_b \gamma_{b'}}\,
   A_{\gamma_{a'} \gamma_{b'} }
                   \left( {\bf k}_a,{\bf k}_b;t_0 \right)\,
   A^*_{\gamma_a \gamma_b } \left({\bf k}_a,{\bf k}_b ;t_0\right)
   \, .
 \label{wf-9}
 \end{equation}
We made the ansatz
$\rho _{\gamma_a \gamma_b,\gamma_{a'} \gamma_{b'} } =
   \rho _{\gamma_a \gamma_{a'}}\,
   \rho _{\gamma_b \gamma_{b'}}$
which factorizes initial density matrix
$\rho _{\gamma_a \gamma_b,\gamma_{a'} \gamma_{b'} }$
in such a way that independent emission of the two particles is ensured.

After straightforward algebra we write expression for the two-particle
probability
\begin{eqnarray}
P_2(q,K) =
&&
\int  d^4 X  S\left( X,K+\frac{q}{2} \right)
\int  d^4 Y  S\left( Y,K-\frac{q}{2} \right) \, \pm \,
\nonumber \\
&&  \pm \,
\int  d^4 X \,  e^{i  q\cdot X } S(X,K)
\int  d^4 Y  \, e^{-i  q\cdot Y } S(Y,K)
\ .
\label{wf-10}
\end{eqnarray}
Finally we get the two-particle correlator $C(q,K)$ (\ref{i2}),
as a ratio of two-particle probability (\ref{wf-10}) and single-particle
probabilities (\ref{wf-4}),
where the source function $S(X,K)$ is defined in accordance with
Eq.~(\ref{i3b}), i.e. all integrations are taken at emission
times or on the freeze-out hyper-surface.

\subsection{ Current parametrization of the source }
\label{sec:rel-current}
Let us consider a single-particle state $\Psi_\gamma$ emitted by the
source which we parametrized by the "source current" $J_\gamma(x)$. Its
propagation to the detector is governed by the Klein-Gordon equation
\begin{equation}
(\partial_\mu \partial^\mu + m^2)\Psi_\gamma (x)=J_\gamma (x)
\, ,
\label{2-1}
\end{equation}
where $\partial_\mu \partial^\mu =\partial_t^2-\vec{\nabla}^2$.
The index $\gamma$ denotes a complete set of 1-particle quantum numbers.
(In a basis of wave packets these could contain the centers ${\bf
X}$ of the wave packets of the particles at their freeze-out times
$t$.)

Single-particle momentum amplitude is defined as projection of the wave
function $\Psi_\gamma(x)$ at "detector time $t=x^0$" on the out-state
$\phi^{\rm out}_{\bf k}$,
\begin{eqnarray}
A_\gamma({\bf k})
=
   \lim_{x^0\to\infty}
   \Big(
   \phi^{\rm out,*}_{\bf k} (x), \,
   \Psi_\gamma(x)
   \Big)
=
   \lim_{t\to\infty}
   \int d^3x\,
   \phi^{\rm out,*}_{\bf k} ({\bf x},t) \,
   i \stackrel{\leftrightarrow}{\partial } _t \,
   \Psi_\gamma({\bf x},t) \, ,
 \label{2-2}
 \end{eqnarray}
where by definition
$a(t) \stackrel{\leftrightarrow}{\partial } _t b(t) \equiv
a\partial_t b-(\partial_t a)b$.
We assume that the detector measures asymptotic momentum
eigenstates, i.e. that it acts by projecting
the emitted single-particle state onto
\begin{equation}
  \phi^{\rm out}_{\bf k}({\bf x},t)
  = e^{ -i\omega ({\bf k})t+i{\bf k}\cdot {\bf x}}
  \equiv f^{(+)}_k(x)
\ ,
\label{2-3}
\end{equation}
where $\omega ({\bf k})=  \sqrt{m^2+{\bf k}^2}$.
Then, momentum amplitude can be rewritten as
%
\begin{eqnarray}
A_\gamma({\bf k}) =
 \lim_{x^0\to\infty}  \int d^3x\, \int d^4y\, f^{(+),*}_k (x) \
  i \stackrel{\leftrightarrow}{\partial } _{x^0} \,
  G_R(x-y)\, J_\gamma (y)
\, .
\label{2-9}
\end{eqnarray}
Substituting to (\ref{2-9}) the Green's function
($f^{(-)}_k(x)\equiv e^{ i\omega ({\bf k})t-i{\bf k}\cdot {\bf x}}$)
$$G_R(x-y)=
 i \, \theta \left( x^0-y^0 \right)
 \int \frac{d^3k}{(2\pi )^3 2\omega ({\bf k})} \,
 \Big[ f^{(+)}_k(x) \,f^{(+),*}_k(y)- f^{(-)}_k(x) \,f^{(-),*}_k(y)
 \Big]\, ,$$
which satisfies equation
$(\partial_\mu \partial^\mu + m^2)G_{R}(x-y)=\delta ^4(x-y)$,
and using orthogonal properties of the basic functions,
$\int d^3x \, f^{(\pm),*}_{k_a}(x) \,
i \stackrel{\leftrightarrow}{\partial }_{x^0} f^{(\pm)}_{k_b}(x)
=
\pm (2\pi )^3\, 2\omega ({\bf k}_a)\, \delta ^3({\bf k}_a-{\bf k}_b)$,
and
$\int d^3x\, f^{(+),*}_{k_a}(x) \,
i \stackrel{\leftrightarrow}{\partial}_{x^0} f^{(-)}_{k_b}(x) = 0$,
after straightforward calculation we come to the answer
\begin{eqnarray}
A_\gamma({\bf k}) =
i \int d^4y \, e^{i\omega ({\bf k})y^0-{\bf k}\cdot {\bf y}}
J_\gamma (y)
\, ,
\label{2-10}
\end{eqnarray}
where integration is taken over infinite space-time volume and that is why
the finiteness in space and time of the particle source which we deal with
is accumulated in the "cut-function" $J_\gamma (y)$.
Moreover, it should be pointed out that amplitude $A_\gamma({\bf k})$ in
(\ref{2-10}) is nothing more as the
on-shell Fourier transformation of the current, hence in this approach
the amplitude to register the particle with certain momentum ${\bf k}$
directly reflects the model of the source which is settled by the
particular definition of the current $J_\gamma (x)$.

\medskip

The single-particle probability is obtained by averaging (\ref{2-10})
and its complex conjugate with the density matrix defining the source.
This density matrix is characterized by a probability distribution for
the single-particle quantum numbers $\gamma $.
We write
 \begin{eqnarray}
 P_1({\bf p}) =
   \sum  _{\gamma , \, \gamma '} \ \rho _{\gamma \gamma '}\,
   A_{\gamma}({\bf p}) \,  A^*_{\gamma'}({\bf p})
 \, ,
 \label{2-12}
 \end{eqnarray}
Inserting (\ref{2-10}) into (\ref{2-12}) and using definition of the source
function (\ref{i4}) after simple algebra we come to result
\begin{equation}
  P_1({\bf k}) =  \int d^4x\, S(x,k)
  \, ,
 \label{2-18}
\end{equation}
which gives the single-particle probability in the same form as we obtained in
Eq.~(\ref{wf-4}) for the wave function approach but in contrast to the wave
function approach the integration in (\ref{2-18}) is taken over infinite
space-time volume.

\medskip


Two-particle momentum amplitude is defined as projection
of the symmetrized (untisymmetrized) two-particle wave function
$\psi_\gamma(x_a,x_b)$ at "detector times
$x_a^0\to\infty $ and $x_b^0\to\infty $"
where the index $\gamma$ denotes a complete set of two-particle quantum
numbers, on the momentum eigen state
$\phi^{\rm out}_{{\bf k}_a,{\bf k}_b}$,
\begin{eqnarray}
A_\gamma({\bf k}_a,{\bf k}_b) &=&
   \lim_{x_a^0\to\infty}\,
   \lim_{x_b^0\to\infty}\,
   \Big(
   \phi^{\rm out,*}_{{\bf k}_a,{\bf k}_b} (x_a,x_b), \,
   \Psi_\gamma(x_a,x_b)
   \Big)
\nonumber \\
&=&
   \lim_{x_a^0\to\infty}\,
   \lim_{x_b^0\to\infty}\,
   \int d^3x_a\, d^3x_b\,
   \phi^{\rm out,*}_{{\bf k}_a,{\bf k}_b} ( x_a, x_b) \,
   i \stackrel{\leftrightarrow}{\partial } _{x_a^0} \,
   i \stackrel{\leftrightarrow}{\partial } _{x_b^0} \,
   \Psi_\gamma( x_a, x_b) \, ,
 \label{2-19}
 \end{eqnarray}
We label out-state by the values of measured momenta, i.e.
${\bf k}_a$ and ${\bf k}_b$. The out-state at detector times
$x_a^0\to\infty \, , \ \ x_b^0\to\infty $
reads
 \begin{eqnarray}
  \phi^{\rm out}_{{\bf k}_a,{\bf k}_b}(x_a,x_b)
  =
  f^{(+)}_{k_a}(x_a) \,  f^{(+)}_{k_b}(x_b) \, .
 \label{2-20}
 \end{eqnarray}

If the source is completely chaotic, i.e. the particles are emitted
independently, that implies that the two-particle wave function is
a product of two single-particle ones.
For pairs of identical bosons (fermions) the two-particle
wave function describing their propagation towards the detector must be
symmetrized (anti-symmetrized).
Taking the same arguments as for the wave function approach about delay of
emission of one particle with respect to another one we write
\begin{eqnarray}
   \Psi_{\gamma_a, \gamma _b}(x_a, x_b)
&=&
   \textstyle{\frac{1}{\sqrt{2} }} \big[
   \Psi_{\gamma _a}(x_a) \,
   \Psi_{\gamma _b}(x_b) \,
   \pm
   \Psi_{\gamma _a}(x_b) \,
   \Psi_{\gamma _b}(x_a) \,
   \big]=
\nonumber \\
&& \hspace{-3.7cm} =
\int d^4y_a \, d^4y_b \, G_R(x_a-y_a) \, G_R(x_b-y_b)\,
  \textstyle{\frac{1}{\sqrt{2} }}
\Big[ J_{\gamma _a}(y_a) \, J_{\gamma _b}(y_b) \,
     \pm
      J_{\gamma _a}(y_b) \, J_{\gamma _b}(y_a) \, \Big]
\nonumber \\
&& \hspace{-3.7cm} =
\int \frac{d^3k_a}{(2\pi )^3 2\omega ({\bf k}_a)}\,
     \frac{d^3k_b}{(2\pi )^3 2\omega ({\bf k}_b)} \,
    f^{(+)}_{k_a}(x_a) \, f^{(+)}_{k_b}(x_b) \,
       \textstyle{\frac{1}{\sqrt{2} }} \Big[
       \tilde{J}_{\gamma _a}(k_a) \, \tilde{J}_{\gamma _b}(k_b) \,
       \pm
       \tilde{J}_{\gamma _a}(k_b) \,  \tilde{J}_{\gamma _b}(k_a) \,
       \Big]
\,,
\label{2-21}
\end{eqnarray}
where we use solution of the Klein-Gordon equation (\ref{2-1}) which
is obtained with a help of the Green's function $G_R(x-y)$ and
\begin{eqnarray}
\tilde{J}_\gamma (k) =
\int d^4y \, e^{i\omega ({\bf k})y^0-{\bf k}\cdot {\bf y}}
J_\gamma (y)
\, ,
\label{2-11a}
\end{eqnarray}
is the on-shell Fourier transformed source current.
We do not write in (\ref{2-21}) the negative-frequency piece of the
Green's function because it evidently disappears on the next step: a
projection of the wave function $\Psi_{\gamma_a, \gamma _b}(x_a, x_b)$
onto out-state.
Indeed, to obtain the momentum amplitude we substitute
expression (\ref{2-21}) into (\ref{2-19}) and use orthogonality relations
of the basic functions $f^{(\pm)}_k(x)$.
All this results in a simple final expression
\begin{equation}
A_{\gamma_a, \gamma _b}({\bf k}_a,{\bf k}_b)
=
  \textstyle{\frac{1}{\sqrt{2} }} \Big[
  \tilde{J}_{\gamma _a}(k_a) \, \tilde{J}_{\gamma _b}(k_b) \,
  \pm
  \tilde{J}_{\gamma _a}(k_b) \,  \tilde{J}_{\gamma _b}(k_a) \,
  \Big]_{k_a^0=\omega ({\bf k}_a), \, k_b^0=\omega ({\bf k}_b)}
\, .
\label{2-23}
\end{equation}
%


The two-particle probability is obtained by averaging amplitude (\ref{2-23})
and its complex conjugate with the density matrix defining the source.
This density matrix is characterized by a probability distribution for
the two-particle quantum numbers $(\gamma_a,\gamma_b)$
\begin{eqnarray}
   P_2({\bf k}_a,{\bf k}_b) =
   \sum  _{\gamma_a \gamma_b,\gamma_{a'} \gamma_{b'} }
   \ \rho _{\gamma_a \gamma_b,\gamma_{a'} \gamma_{b'} }\,
   A_{\gamma_a \gamma_b } \left({\bf k}_a,{\bf k}_b \right) \,
   A^*_{\gamma_{a'} \gamma_{b'} } \left({\bf k}_a,{\bf k}_b \right)
\, ,
\label{2-24}
\end{eqnarray}
As in the wave function approach we made the ansatz
 $  \rho _{\gamma_a \gamma_b,\gamma_{a'} \gamma_{b'} } =
   \rho _{\gamma_a \gamma_{a'}}\,
   \rho _{\gamma_b \gamma_{b'}}$,
which factorizes in such a way that independent emission of the
two particles is ensured.

Substituting momentum amplitude (\ref{2-23}) into (\ref{2-24}),
using definition (\ref{i4}) of the source function $S(Y,K)$,
we can write for the two-particle probability
\begin{equation}
   P_2({\bf k}_a,{\bf k}_b) =
   \int d^4X\, S(X,k_a)
   \int d^4Y\, S(Y,k_b)
   \pm
   \int d^4X\,
   e^{i q\cdot X}\,  S(X,K)
   \int d^4Y\,
   e^{-i q\cdot Y}\, S(Y,K)
\, ,
\label{2-28}
\end{equation}
which coincide with expression (\ref{wf-10}) obtained in wave
function approach, consequently, we obtain correlator in the form
(\ref{i2}).
The integration on the right hand side of Eq.~(\ref{2-28}) is just
taken over an infinite space-time interval, whereas in (\ref{wf-10})
the integration is taken over freeze-out hyper surface, or over
initial times.
%

\subsection{Wave function  parametrization versus current
            parametrization.
            Nonrelativistic approach}
\label{sec:nonrel}
The goal of this subsection is to put in correspondence the ``wave
function'' approach which was elaborated in the Section II.A to the
``current'' approach of the Section II.B.
To do this we consider
this correspondence first in the non-relativistic limit and then
fully relativistic comparing will be carried out.

We are going to obtain the current approach in the
non-relativistic limit (see Appendix \ref{append:nonrelat}).
We make a standard unitary transformation of the wave function to extract
oscillations associated with particle mass
$\Psi_\gamma (x) = e^{-imx^0}\psi_\gamma (x)$.
With respect to the new wave function $\psi_\gamma (x)$ the basic
equation (\ref{2-1}) reads
\begin{equation}
\left( i\partial_t + \frac{1}{2m} \, \nabla^2 \right)
\psi_\gamma(t,{\bf x}) = j_\gamma (t,{\bf x})
\, ,
\label{4:5a}
\end{equation}
where
\begin{equation}
j_\gamma (t,{\bf x})=- \frac{1}{2m} \, e^{imt} J_\gamma (t,{\bf x})
\, .
\label{cor:6}
\end{equation}
and we skipped all terms of the order $1/c^2$ and higher, they
serve as relativistic corrections.
By this derivation we put in correspondence the current in the
relativistic parametrization of the source with the current in the
non-relativistic one.

With respect to the quantum state $\psi_\gamma(t,{\bf x})$ the
momentum amplitude can be rewritten in the following way
\begin{eqnarray}
A_\gamma({\bf k})
= \!
\lim_{x^0\to\infty} \int d^3x \! \int d^4y f^{(+),*}_k (x)
 G_0(x-y)\, j_\gamma (y) \!
= \!
i\int d^4y \, e^{i\omega ({\bf k})y^0-{\bf k}\cdot {\bf y}}
j_\gamma (y)
=
i\, \tilde{j}_\gamma (k)
 ,
\label{2-9n}
\end{eqnarray}
where
\begin{equation}
G_0(x-y)
 = - \int \frac{d^4k}{(2\pi )^4}\,
   \frac{\displaystyle e^{-i k\cdot (x-y)}}{k^0-\omega ({\bf k})+i\epsilon }
=i \, \theta (x^0-y^0) \int \frac{d^3k}{(2\pi )^3}
  e^{-i k\cdot (x-y)} \Big|_{ k^0=\omega ({\bf k})}
\,,
\label{cor:11}
\end{equation}
is the Green's function which satisfies equation
$ \left( i\partial_t + \nabla^2 /2m\right)G_0(x-y)=\delta ^4(x-y)$
and $\tilde{j}_\gamma (k)$ is the on-shell Fourier transformation of
the current.

Compare the amplitude (\ref{2-9n}), i.e.
$A_\gamma({\bf k})=i\, \tilde{j}_\gamma (k)$, with the correspondent
amplitude obtained in the relativistic case (\ref{2-10}) we see
that they coincide with one another, just in place of the capital
letter $J$ one should put a small one.
Hence, the same transformation should be done in the definition of the
source function (\ref{i4}).

Non-relativistic Schr\"odinger equation
$ \left( i\partial_t - \hat{H}({\bf x}) \right) \psi_\gamma(x)=0 $
supplemented by the initial condition,
$\psi_\gamma(t=t_0,{\bf x})=\Phi_\gamma({\bf x})$,
can be written, as was shown in Appendix \ref{append:initial}
(the generalized Cauchy problem \cite{vladimirov}),
in the form of the Schr\"odinger equation with the source on the
r.h.s. of equation which is defined at initial time $t=t_0$,
\begin{equation}
\left(  i\partial_t - \hat{H}({\bf x}) \right) \psi_\gamma(t,{\bf x})
= i \, \Phi_\gamma({\bf x}) \, \delta(t-t_0)
\, ,
\label{cor:8}
\end{equation}
which is valid for $t\ge t_0$.
Then, one can solve this equation with the help of the Green's
function (\ref{cor:11}) and write solution in the following form
\begin{eqnarray}
\psi_\gamma(t-t_0,{\bf x})
&=&
i \, \int  d^3y \,
G_0(t-t_0,{\bf x}- {\bf y}) \, \Phi_\gamma({\bf y})
\nonumber \\
&=&
\theta(t-t_0) \,
\int \frac{d^3p}{(2\pi)^3} \, d^3x' \,
e^{-i \omega({\bf p})(t-t_0) + i{\bf p}\cdot ({\bf x}- {\bf x}') }
\psi_\gamma(t_0,{\bf x}')
\, .
\label{cor:15}
\end{eqnarray}
As a matter of fact, this solution coincides with that one obtained
with the help of the evolution operator,
$ \psi_\gamma(t-t_0,{\bf x})=
\theta(t-t_0) \, e^{-i \hat{H}(\hat{{\bf p}},{\bf x})(t-t_0) }
\psi_\gamma(t_0,{\bf x})$, which we exploited in the wave function
approach in paragraph \ref{sec2-1}.

On the other hand, solving the Cauchy problem in this way one can
consider the expression on the r.h.s. of eq.(\ref{cor:8}) as
a specific current
\begin{equation}
j_\gamma(t,{\bf x})= i \, \psi_\gamma(t,{\bf x}) \, \delta(t-t_0)
\, .
\label{cor:9}
\end{equation}
Then, going through all preceding steps to evaluate the source
function $S(X,K)$ with making use of this specific current
one evidently discovers that the source functions (\ref{i4})
and (\ref{i3b}) coincide with one another.
We regard this result as first example when two types of
parametrization of the source can give the same answer for specific
connection between current and wave function given at freeze-out
times.

We are going now to prove that the same is valid in more general case.
Indeed, let us write once more the solution of eq.(\ref{4:5a})
\begin{eqnarray}
\psi_\gamma(t,{\bf x})
&=&
\int  d^4z \, G_0(t-z_0,{\bf x}- {\bf z}) \, j_\gamma (z_0,{\bf z})
\nonumber \\
&=&
i \, \int d^4z \,  d^3y \,
G_0(t-t_0,{\bf x}- {\bf y}) \, G_0(t_0-z_0,{\bf y}- {\bf z})
  j_\gamma (z_0,{\bf z})
\, ,
\label{cor:16d}
\end{eqnarray}
where in the second line we split the Green's function at the point
$t=t_0$ using the group property of the Green's functions.
We are making now the physical input: let us prepare the initial
state, which will be used in the wave function parametrization of the
source,
$\Phi_\gamma({\bf y})=\psi_\gamma(t_0,{\bf y})$, in the following way
(note, up to now we did not specialize a generation of the wave
function at freeze-out times)
\begin{equation}
\Phi_\gamma({\bf y})
= \int  d^4z \, G_0(t_0-z_0,{\bf y}- {\bf z}) j_\gamma (z_0,{\bf z})
\, .
\label{cor:16c}
\end{equation}
Then, rewriting the second line in (\ref{cor:16d}) with making use of
the state $\Phi_\gamma({\bf y})$ (just defined in (\ref{cor:16c}))
we obtain the single-particle quantum state,
$\psi_\gamma(x)$, at the times which are after $t=t_0$, i.e. after
freeze-out, in the following form
\begin{equation}
\psi_\gamma(t,{\bf x})
=
i \, \int  d^3y \,
G_0(t-t_0,{\bf x}- {\bf y}) \, \psi_\gamma(t_0,{\bf y})
\, .
\label{cor:16a}
\end{equation}
What is most interesting, expression (\ref{cor:16a}) is exactly a
solution of the Schr\"odinger equation (\ref{cor:8}) with
$\psi_\gamma(t_0,{\bf y})$ as the initial condition
(see (\ref{cor:15})).

Let us make one note.
In eq.(\ref{cor:16d}) after splitting of the Green's function we meet
the product of two $\theta$-functions, $\theta(t-t_0)\theta(t_0-z_0)$.
Because we are interesting in detector times the value of time $t$
goes to infinity and we can avoid the first $\theta$-function.
At the same time the second $\theta$-function cuts an action of the
source current at the times $t=t_0$.
But this feature does not distort the influence of the current if
$t_0$ is the freeze-out time.
In other words, we assume that a life time of the current coincides
with a life time of the fireball,
$j_\gamma (t,{\bf x})\propto \theta(t_0-t)$.

So, we obtain the same quantum state
$\psi_\gamma(t,{\bf x})$ in two approaches:
1) The current parametrization of the source, expression
(\ref{cor:16d}), first line, which is solution of eq.(\ref{4:5a}); and
2) The wave function parametrization
of the source, expression (\ref{cor:16a}), which is solution of the
Cauchy problem for the Schr\"odinger equation.
If we start a description of the propagation of the particle to
detector from the quantum state $\psi_\gamma(t,{\bf x})$ taken
from (\ref{cor:16a}) and go through all steps to the source function
$S(X,K)$ we come to expression (\ref{i3b}) which is evaluated
with the help of the initial states $\psi_\gamma(t_0,{\bf y})$.
In fact, as was shown in the section \ref{sec2-1}, by this we obtain
the source function exploiting the
wave function parametrization of the source.
On the other hand, we can start the evaluation from the same quantum
state $\psi_\gamma(t,{\bf x})$ taking it in the form (\ref{cor:16d})
(first line).
Then, we come to the source function (\ref{i4}) obtained in the
current approach.
Meanwhile, the starting point for both expressions is the same state
$\psi_\gamma(t,{\bf x})$ (what gives the same amplitude, the same
single-particle probability and so on).
Hence, if the initial wave function, $\Phi_\gamma({\bf x})$, and the
current, $j_\gamma(x)$, are connected to one another by
eq.(\ref{cor:16c}), then both evaluations of the source function
$S(X,K)$ give the same result.
Thus, the single-particle Wigner functions constructed in both
approaches are equal
 \begin{eqnarray}
S(Y,K)
&=& \delta(Y^0-t_0) \int d^4y\, \delta(y^0) \, e^{i K\cdot y}\,
  \sum_{\gamma ,\, \gamma '} \rho_{\gamma \gamma '}\,
  \psi_\gamma \left(Y+{\textstyle{y\over 2}}\right) \,
  \psi_{\gamma '}^*\left(Y-{\textstyle{y\over 2}}\right)
\nonumber \\
&=&
\int d^4y\,  e^{i K\cdot y}\,
  \sum_{\gamma ,\, \gamma '} \rho_{\gamma \gamma '}\,
  j_\gamma \left(Y+{\textstyle{y\over 2}}\right) \,
  j_{\gamma '}^*\left(Y-{\textstyle{y\over 2}}\right)
\, .
 \label{cor:16e}
 \end{eqnarray}
Consequently, the single-particle spectrum and two-particle
correlations taken in the wave function parametrization of the source
coincide with the respective spectra taken in the current
parametrization.

Connection (\ref{cor:16c}) between current and wave function at
freeze-out times has a transparent physical interpretation:
the action of the current which describes in a semi-classical way a
creation of
secondary particles during the life time of the fireball can be
accumulated in the wave function at freeze-out times.
That is why, it does not matter what quantity is used then to describe
the free propagation of the particles to detector.
Moreover, the correspondence (\ref{cor:16a}), as it is seen in
Fig.\ref{figure:const-t}, results in extension of
the effective volume where initial wave function is given by adding
a spherical layer for the radii $r$ in the limits $R_0 \le r \le R_1$.
Indeed, all particles which were emitted during life time of the
fireball from the boundary $F_1R_0$ accumulated now on the space-like
segment $F_1F_2$.
This means that the wave function $\psi_\gamma(t_0,{\bf x})$ given on
the extended space-like hypersurface, segment $F_0F_1F_2$,
takes into account all secondary particles which were ``produced'' by
the source current.

\subsection{ Wave-function versus current parametrization
of the source in relativistic approach }

\begin{figure}
{\includegraphics[width=11cm, height=8cm, angle=0]{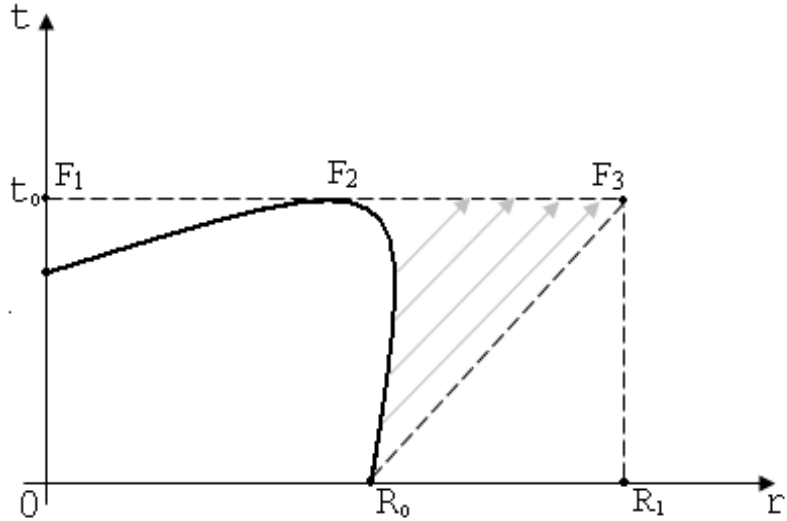} }
\caption{
A sketch of the freeze-out hyper-surface (solid curve) for a spherically
symmetric expansion of the fireball.}
\label{figure:freeze-out}
\end{figure}

First, we consider the wave-function parametrization of the source.
Single-particle momentum amplitude is evaluated as projection of the
wave function, $\Psi_\gamma(t,{\bf x})$, taken at asymptotic times
onto out-state $\phi^{\rm out}_{\bf p}$
\begin{eqnarray}
A_\gamma({\bf p})
=
\lim_{t\to\infty} \int d^3x\, \phi^{\rm out,*}_{\bf p} (t,{\bf x}) \,
   i \stackrel{\leftrightarrow}{\partial } _t \,
   \Psi_\gamma(t,{\bf x})
\, ,
\label{cor:41}
\end{eqnarray}
where $\phi^{\rm out}_{\bf p}(t,{\bf x})
        = e^{-i\, \omega({\bf p}) t+ i{\bf p}\cdot {\bf x} }$.
The wave function, $\Psi_\gamma(t,{\bf x})$ is solution of the
Klein-Gordon equation,
$(\partial_\mu \partial^\mu + m^2)\Psi_\gamma (x)=0$, which is
supplemented by the initial conditions
\begin{equation}
\Psi_\gamma(t,{\bf x})\big|_{\,t=t_0}=\Phi_{\gamma 0}({\bf x})\, ,
\ \ \ \ {\rm and} \ \ \ \
\frac{\partial \Psi_\gamma(t,{\bf x})}{\partial t}\bigg|_{\,t=t_0}
= \Phi_{\gamma 1}({\bf x})
\, .
\label{rel:20}
\end{equation}
As we show in Appendix \ref{append:initial} (see (\ref{c:24})) this
problem can be formulated as equation with a source which constructed
with a use of the initial conditions (\ref{rel:20})
\begin{equation}
(\partial_\mu \partial^\mu + m^2)\Psi_\gamma(t,{\bf x})
= \Phi_{\gamma 1}({\bf x}) \, \delta(t-t_0) + \Phi_{\gamma 0}({\bf x}) \,
  \delta'(t-t_0)
\, ,
\label{rel:24}
\end{equation}
which is valid for times $t\ge t_0$.
Solving this equation with a help of the Green's function and
inserting solution to (\ref{cor:41}) one can write the amplitude in the
following form
\begin{eqnarray}
A_\gamma({\bf p}) =
\lim_{x^0\to\infty} \int d^3x \,  \int d^4y \,  \delta(y^0-t_0)\,
f^{(+),*}_p (x) \, i
\frac{\stackrel{\leftrightarrow}{\partial } }{\partial x^0 } \,
   G_R(x-y) \,
\frac{\stackrel{\leftrightarrow}{\partial } }{\partial y^0} \,
   \Psi_\gamma(y)
\, ,
\label{cor:41a}
\end{eqnarray}
where $f^{(+)}_p(x)= \phi^{\rm out}_{\bf p}(t,{\bf x})$.
Taking $G_R(x-y)$ in the explicit form and using the orthogonal
properties of the basic functions $f^{(\pm)}_p (x)$ we come to the
answer
\begin{eqnarray}
A_\gamma({\bf p})
=
 \int \, d^4y \,  \delta(y^0-t_0)\, \left[
f^{(+),*}_p (y) \, i \frac{\stackrel{\leftrightarrow}{\partial } }
  {\partial y^0} \, \Psi_\gamma(y) \right]
=
\Psi^{(+)}_\gamma(t_0,{\bf p})
\, ,
\label{cor:42}
\end{eqnarray}
where $\Psi^{(+)}_\gamma(t_0,{\bf p})$ is the Fourier component of a
positive-energy piece of the function $\Psi_\gamma(t_0,{\bf y})$.
Note, the wave function consists from two contributions, positive- and
negative-energy defined,
$\Psi_{\gamma}(x)=\Psi^{(+)}_{\gamma}(x)+\Psi^{(-)}_{\gamma}(x)$,
respectively.
Actually, for the sake of simplicity we consider a flat space-like
hypersurface, $t=t_0$, the segment $F_1F_2F_3$, as it is depicted in
Fig.\ref{figure:freeze-out}.
%
At the same time, the amplitude (\ref{cor:42}) can be expressed in the
covariant form  as well
\begin{eqnarray}
A_\gamma({\bf p})
=
 \int \, d\sigma^\mu(y) \,
f^{(+),*}_p (y) \, i \frac{\stackrel{\leftrightarrow}{\partial } }
  {\partial y^\mu} \, \Psi_\gamma(y)
\, .
\label{cor:43}
\end{eqnarray}
where $y^0=\sigma({\bf y})$ is the space-like hypersurface on which the
wave function and its derivative are given.

Formula (\ref{cor:42}) gives a parametrization of the
probability amplitude by the initial values of the wave function and
its derivative at freeze-out times.
The next steps are the same as in the section \ref{sec:rel-current}.
The single-particle probability is obtained with a help of the density
matrix:
$ P_1({\bf p})=\sum  _{\gamma , \, \gamma '}\rho _{\gamma \gamma '}\,
  A^*_{\gamma } ( {\bf p} ) \, A_{\gamma '} ( {\bf p} )
  =\sum  _{\gamma , \, \gamma '}\rho _{\gamma \gamma '}\,
  \Psi^{(+),*}_\gamma(t_0,{\bf p})\Psi^{(+)}_{\gamma'}(t_0,{\bf p})$.
If we now represent the functions $\Psi^{(+)}_{\gamma}(t_0,{\bf p})$
as the Fourier integral
\begin{equation}
\Psi^{(+)}_{\gamma}(t_0,{\bf p})
=
\int d^3x\,e^{-i{\bf p}\cdot {\bf x}} \, \Psi^{(+)}_{\gamma}(t_0,{\bf x})
=
\int d^4x \, \delta(x^0-t_0)\, e^{ip\cdot x}\, \Psi^{(+)}_{\gamma}(x)
\, ,
\label{relwf:2}
\end{equation}
and insert it to $P_1({\bf p})$ we come to the standard expression of
the single-particle probability
$P_1({\bf k}) =  \int d^4X\, S(X,k)$, where we define the source
function
 \begin{equation}
S(X,K)
= \delta(X^0-t_0)
  \int d^4x\, \delta(x^0) \, e^{i K\cdot x}\,
  \sum_{\gamma ,\, \gamma '} \rho_{\gamma \gamma '}\,
  \Psi^{(+),*}_\gamma \left(X+{\textstyle{x\over 2}}\right) \,
  \Psi^{(+)}_{\gamma '}\left(X-{\textstyle{x\over 2}}\right)
\, .
 \label{relwf:3}
 \end{equation}

So, after definition the source function (\ref{relwf:3}), which is
constructed with the use
of the wave function and and its derivative at freeze-out times,
we are ready to compare this parametrization of the source with the
current one.

First of all let us mention that both parametrizations evidently
coincide when one takes the special form of the current,
$J_\gamma(t,{\bf x})=\partial \Psi_\gamma(t,{\bf x})/\partial t \,
\delta(t-t_0) + \Psi_\gamma(t_0,{\bf x}) \, \delta'(t-t_0)$,
which is expression on the r.h.s. of eq.(\ref{rel:24}).
Then, we come to eq.(\ref{2-1}) which is starting point in the current
parametrization of the source.
We show now that the same is valid in more general case (our
consideration is very close to that one developed in the section
\ref{sec:nonrel}).

The single-particle amplitude at asymptotic times (\ref{2-9}) which we
obtained in the section \ref{sec:rel-current} can be written in the
following way
\begin{eqnarray}
\tilde{A}_\gamma({\bf k}) =
  \lim_{x^0\to\infty} \int d^3x\, \int d^4y\, \theta(x^0-y^0) \,
  \left[ f^{(+),*}_k (x) \,
  i  \frac{ \stackrel{\leftrightarrow}{\partial } }{\partial x^0} \,
  G^{(+)}(x-y)\right] \, J_\gamma (y)
\, ,
\label{relwf:4}
\end{eqnarray}
where we use representation (\ref{c:29}) of the retarded Green's
function,
$G_R(x-y)=\theta \left( x^0-y^0 \right) \left[ G^{(+)}(x-y)
- G^{(-)}(x-y) \right]$, and orthogonality relation of the basic set
of functions $f^{(+)}_k (x)$ and $f^{(-)}_k (x)$.
To distinguish the ``current'' amplitude (\ref{relwf:4}) from
the ``wave function'' one (\ref{cor:42}) we marked it by tilde.
On the next step we use the group property of the functions
$G^{(+)}(x-y)$ (see (\ref{c:31}))
\begin{equation}
G^{(+)}(x-y) =
\int  d^4z \, \delta\left(z^0-t_0\right)
\left[  G^{(+)}(x-z)
\frac{ \stackrel{\leftrightarrow}{\partial } }{\partial z^0} \,
        G^{(+)}(z-y) \right]
 \, .
\label{relwf:5}
\end{equation}
Inserting this expression into (\ref{relwf:4}) and making the
following convolution
\begin{equation}
\int d^3x \, f^{(+),*}_k (x) \,i \,
\frac{ \stackrel{\leftrightarrow}{\partial } }{\partial x^0}
\, G^{(+)}(x-z)
=
i\, f^{(+),*}_k (z)
 \, ,
\label{relwf:6}
\end{equation}
we get the amplitude (\ref{relwf:4}) in the form
\begin{eqnarray}
\tilde{A}_\gamma({\bf k})
&=&
\int d^4y\,  d^4z\, \, \delta\left(z^0-t_0\right) \,
  \left[ f^{(+),*}_k (z) \,
  i  \frac{ \stackrel{\leftrightarrow}{\partial } }{\partial z^0} \,
  G^{(+)}(z-y)\right] \, J_\gamma (y)
\nonumber \\
&=&
\int d^4y\,  d^4z\, \, \delta\left(z^0-t_0\right) \,
  \left[ f^{(+),*}_k (z) \,
  i  \frac{ \stackrel{\leftrightarrow}{\partial } }{\partial z^0} \,
  G_R(z-y)\right] \, J_\gamma (y)
\, ,
\label{relwf:7}
\end{eqnarray}
where the last line in (\ref{relwf:7}) is obtained under assumption
that the source current ``works'' just during the life time of the
fireball, i.e. $J_\gamma (t,{\bf x}) \propto \theta(t_0-t)$.
Taking into account this feature one can define the wave function at
freeze-out times as:
\begin{equation}
\Psi_\gamma(t_0,{\bf z})
=
\int d^4y \, G_R(t_0-y^0,{\bf z}-{\bf y}) \, J_\gamma (y^0,{\bf y})
 \, .
\label{relwf:8}
\end{equation}
Inserting this notation to the second line on the r.h.s. of
(\ref{relwf:7}) one can rewrite the amplitude
$\tilde{A}_\gamma({\bf k})$ in the following way
\begin{eqnarray}
\tilde{A}_\gamma({\bf k})
=
\int d^4z\, \, \delta\left(z^0-t_0\right) \,
  \left[ f^{(+),*}_k (z) \,
  i  \frac{ \stackrel{\leftrightarrow}{\partial } }{\partial z^0} \,
  \Psi_\gamma(z)\right]
\, ,
\label{relwf:9}
\end{eqnarray}
Then, as will readily be observed the last expression coincide
literally with the amplitude obtained in the wave function
parametrization of the source (\ref{cor:42}).
Hence, we can write
\begin{equation}
\tilde{A}_\gamma({\bf k})
= A_\gamma({\bf k})
 \, .
\label{relwf:10}
\end{equation}
Because, the group property can be written in covariant form as well
$G^{(+)}(x-y)=
\int_{(\sigma)}  d\sigma^\mu(z) \,G^{(+)}(x-z) \,
\Big(\stackrel{\leftrightarrow}{\partial}/\partial z^\mu \Big)\,
G^{(+)}(z-y)$, one can obtain the amplitude in the covariant form
(\ref{cor:43}).
Then, equality (\ref{relwf:10}) is valid for an arbitrary space-like
hyper-surface.

So, if we keep relation between current and wave function at
freeze-out times in the form (\ref{relwf:8}) we guarantee that the
single-particle momentum amplitude will be the same in both
approaches.
The statement is valid also for two-particle momentum amplitude.
Because the amplitude is the main constructive element of the
single-particle probability (\ref{2-12}) and two-particle probability
(\ref{2-24}), the equality of the amplitudes results in the equality
of probabilities.
This means that the source functions obtained in the wave function
parametrization (\ref{relwf:3}) and in the current parametrization of
the source (\ref{i4}) are equal as well when eq.(\ref{relwf:8}) is
valid.

\begin{figure}
{\includegraphics[width=5cm, height=8cm, angle=-90]{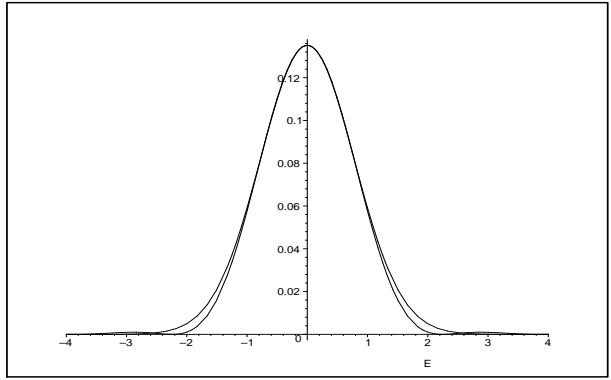} }
\caption{
Comparison of the Fourier coefficients squared of the Gaussian and
$\theta-$function time cutting of a source current.}
\label{figure:gaussian}
\end{figure}

It is necessary to clarify the time structure of the
current $J_\gamma (t,{\bf x})$.
As a source of the single-particle state $\Psi_\gamma (t,{\bf x})$
the current acts during the life time of the fireball or when its
time argument
$t$ is less than freeze-out times, $t \le t_\sigma$:
\begin{equation}
J_\gamma (t,{\bf x}) \propto  \theta( t_\sigma-|t|)
\, .
\label{cor:17}
\end{equation}
In the applications the cutting of the time interval is usually made
in a soft way with a help of the Gaussian function,
$J_\gamma (t,{\bf x}) \propto \exp{(-t^2/2\tau^2)}$,
where $\tau$ is of the same order as $t_\sigma$.
Our previous consideration was based on the rapid cutting of the
current on the freeze-out hyper-surface like that in (\ref{cor:17}).
Can a smooth switching off destroy our scheme?
It is necessary to point out that just the Fourier transformed
quantities enter the single-particle and two-particle probabilities.
Let us look at the shape of the Fourier transformed cutting profiles.
It is interesting to note that the Fourier components of the both time
cutting functions, the Gaussian function and $\theta$-function
(\ref{cor:17}), give
approximately the same bell like dependence on energy variable $E$,
$J(E)=\int^\infty_{-\infty}dt\, J(t)\, \exp{(iEt)}$.
These functions squared, $J(E)^2$, are depicted in
Fig.\ref{figure:gaussian}.
Only a slight difference between these functions is seen and,
therefore, the choice of the type of time cutting function does not
affect much, at least qualitatively they give the same result.
That is why, if we exploit the $\theta$-function cutting rule
(\ref{cor:17}) we obtain the same probabilities to register the
particles as in the case of the Gaussian profile.
%

\section{Discussion and conclusions }

We considered two types of a semi-classical parametrization of the
source which give a transparent scheme of evaluation of the
single-particle spectrum and two-particle correlations:
the wave function and current parametrization of the source.
The main ingredients of the wave function parametrization are the
values of the wave function which are given on freeze-out
hyper-surface (in relativistic approach the values of the wave
function derivative should be given as well).
In describing a propagation of the particles to detector after
freeze-out these values serve as the initial conditions in the Cauchy
problem: all information about
evolution of the fireball is accumulated in the single-particle
wave function, $\psi_\gamma(t_0,{\bf x})$, given at freeze-out times
(see Fig.\ref{figure:freeze-out}).
In relativistic case it is $\Psi_\gamma(t_0,{\bf x})$ and
its derivative $\partial \Psi_\gamma(t_0,{\bf x})/ \partial t$,
then, the relativistic projection onto the out-state results
that just the positive-energy defined part of the
wave-function, $\Psi^{(+)}_\gamma(t_0,{\bf x})$, is exploited.
For the sake of simplicity we discuss here a flat space-like
hyper-surface $t=t_0=$const.
An arbitrary freeze-out hyper-surface is also considered in the paper.

Once the wave function at freeze-out times, $t=t_0$, is given, then,
the single-particle spectrum and two-particle correlations can be
constructed with the help of the single-particle Wigner density
$S(X,K)$ which reads
\begin{equation}
S(X,K)
=
\delta(X^0-t_0) \int d^4x\, \delta(x^0) \, e^{i K\cdot x}\,
  \sum_{\gamma ,\, \gamma '} \rho_{\gamma \gamma '}\,
  \psi_\gamma^* \left(X+{\textstyle{x\over 2}}\right) \,
  \psi_{\gamma '}\left(X-{\textstyle{x\over 2}}\right)
\, ,
\label{d:2}
\end{equation}
where the measure of integration appears as a result of the
transformation:
$d^4x_1\, d^4x_2\, \delta(x_1^0-t_0)\delta(x_2^0-t_0)=
d^4X\,d^4x\, \delta(X^0-t_0) \, \delta(x^0)$ with
$X=(x_1+x_2)/2, \ x=(x_1-x_2)$.
To obtain the source function in relativistic picture one should put
in (\ref{d:2}) the functions $\Psi^{(+)}_\gamma(x)$ in place of
$\psi_\gamma(x)$, then, we come to expression (\ref{relwf:3}).

We propose a scheme to generate the values of the wave function at
freeze-out times.
This can be done with a help of the current which parameterizes the
source
\begin{equation}
\Psi_\gamma(t_0,{\bf x})
=
\int dy^0\, d^3y \, G_R(t_0-y^0,{\bf x}-{\bf y}) \,
  J_\gamma (y^0,{\bf y})
 \, ,
\label{d:3}
\end{equation}
where $J_\gamma (y^0,{\bf y}) \propto \theta(t_0-y^0)$, i.e. the life
time of the current equals the life time of the fireball.
If the wave function $\Psi_\gamma(t_0,{\bf x})$ and the current
$J_\gamma (y^0,{\bf y})$ which parameterizes the source are in relation
(\ref{d:3}), then, the single- and two-particle spectra evaluated in
the wave function parametrization are equal to the same quantities in the
current parametrization.
This results in equality of the source function (\ref{d:2})
(or (\ref{relwf:3}) in relativistic approach) obtained in the wave
function parametrization of the source with the source function
(\ref{i4}) obtained in the current parametrization.
Moreover, the correspondence (\ref{d:3}) results in extension of the
space-like piece of the freeze-out hyper-surface by including the
new piece which
is created by the particles emitted from the fireball during its life,
$0\le t \le t_0$, through the time-like part of the freeze-out
hyper-surface.
As it is seen in Fig.\ref{figure:const-t} the space-like part of the
freeze-out hyper-surface, segment $F_0F_1$, is extended by adding a
new piece, segment $F_1F_2$, which is created by the particles emitted
from the boundary, segment $F_1R_0$.
The same picture takes the place for the freeze-out hyper-surface of
an arbitrary shape which is sketched in Fig.\ref{figure:freeze-out}:
the segment $F_2F_3$ accumulates the particles emitted from the
time-like part of the freeze-out hyper-surface during time span $t_0$.
Hence, if the wave function at freeze-out times is generated by the
source current as in (\ref{d:3}), then, the wave function accumulates
information about all particles emitted from the fireball.
Moreover, the correspondence (\ref{d:3}) (the correspondence
(\ref{cor:16a}) in nonrelativistic case), as it is seen in
Figs.\ref{figure:const-t}, \ref{figure:freeze-out} results in
extension of the size of the fireball: the wave function
parametrization
reflects the radius of the system which is $R_1$ (but not $R_0$).
So, effectively the size of the system is bigger, an extended
volume
includes the spherical layer, $R_0 \le r \le R_1$, which contains
free particles emitted from the boundary of the fireball.

However, there is a source of particles, for instance pions, which
creates particles after freeze-out, for instance a decay of long
lived resonances.
It can be formalized by introducing of a ``post freeze-out'' current,
$J_{\rm pfo}(x)$.
Basically, from the very beginning we can separate current into two
parts: the first part is a
current before freeze-out, $I_\gamma(x)$, and the second part is a
current after freeze-out, $I_{\rm pfo}(x)$,
$$J_\gamma(t,{\bf x})=\theta(t_0-t)\, I_\gamma(t,{\bf x})
     +\theta(t-t_0)\, I_{\rm pfo}(t,{\bf x}).$$
Then, the momentum amplitude (\ref{relwf:9}) should be modified to
the following form
\begin{eqnarray}
A_\gamma({\bf p})
=
 \int \, d^4x \,  \delta(x^0-t_0)\,
f^{(+),*}_p (x) \, i \frac{\stackrel{\leftrightarrow}{\partial } }
  {\partial x^0} \, \Psi_\gamma(x)
+
i\, \int d^4x \, \theta(x^0-t_0) \, f^{(+),*}_p (x)\,
I_{\rm pfo}(x)
\, ,
\label{d:4}
\end{eqnarray}
where the second term on the r.h.s. of equation reflects the sources
of particles which appear after freeze-out, i.e. for times $t \ge t_0$.
It turns out that these particles give contribution just into the
single-particle spectrum.
Because for the particles which are created after freeze-out a
symmetrization, for instance of two-particle wave function,
starts when the particles are separated by big distances the
two-particle momentum probability, $P_2(K,q)$, has appreciable values
just for small relative momenta, $|\,{\bf q}\,| \le 10\div 20$~MeV/c.
These values of the relative momenta are not experimentally
``visible''.

\medskip

{\bf Acknowledgements:} The work of D. Anchishkin was partially
supported by the CERN TH Division (The Scientific Associates
programme) and by the program "Fundamental properties of the
physical systems under extreme conditions" of the Section for
physics and astronomy of the National Academy of Sciences of
Ukraine.
The work of U. Heinz was supported by the U.S. Department of
Energy under contract DE-FG02-01ER41190.


\appendix

\section{ Nonrelativistic approximation of the Klein-Gordon equation }
\label{append:nonrelat}
In this appendix we are going to obtain a relation between the current
which appears as a right part (source) of the Klein-Gordon equation
and a current (source) of the Schr\"odinger equation.
For this purpose we take eq. (\ref{2-1}), which determines a current
parametrization of the source, and consider it in a non-relativistic
limit.
This equation can be rewritten in the following way
\begin{equation}
(\partial_\mu \partial^\mu + m^2)\Psi_\gamma (x) =
-(i\partial_t + \sqrt{m^2-\nabla^2})\, (i\partial_t - \sqrt{m^2-\nabla^2})
\Psi_\gamma (t,{\bf x}) = J_\gamma (t,{\bf x})
\, ,
\label{4:1}
\end{equation}
where $t=x^0$.
We make a standard unitary transformation of the wave function to extract
oscillations associated with particle mass
\begin{equation}
\Psi_\gamma (x) = e^{-imx^0}\psi_\gamma (x)
\, .
\label{4:2}
\end{equation}
With respect to the new wave function $\psi_\gamma (x)$ eq.~(\ref{4:1})
reads
\begin{equation}
(i\partial_t + m + \sqrt{m^2-\nabla^2})\,
(i\partial_t + m - \sqrt{m^2-\nabla^2})
\psi_\gamma (t,{\bf x}) = - e^{imt} J_\gamma (t,{\bf x})
\, .
\label{4:3}
\end{equation}
It is necessary to point out that from now on the energy operator
$i\partial_t$ with respect to the wave function $\psi_\gamma (x)$ is
an operator of the kinetic energy because
$i\partial_t \psi_\gamma (t,{\bf x})=e^{imt}\left( i\partial_t-m\right)
\Psi_\gamma (t,{\bf x}) $.
We shall also quote our consideration to positive enegies.
That is why in the non-relativistic approximation we have the
following inequalities for energy and momentum operators:
$\langle i\hbar\partial_t \rangle /mc^2 \ll 1$ and
$\langle -i\hbar\nabla \rangle /mc \ll 1$, where the broken brackets
mean an averaging over some single-particle quantum state.
Hence, the operator in the first bracket on the l.h.s. of eq.~(\ref{4:3})
is a positive definite operator (it does not have zero eigenvalues).
As a consequence, it always has inverse operator, that is why we write
\begin{equation}
\left( i\partial_t + m - \sqrt{m^2-\nabla^2}\right)
\psi_\gamma (t,{\bf x}) =
- \left( i\partial_t + m + \sqrt{m^2-\nabla^2}\right) ^{-1}\,
e^{imt} J_\gamma (t,{\bf x})
\, .
\label{4:4}
\end{equation}
With making use of the relation
$-i\nabla \psi_\gamma (t,{\bf x})  \ll m\psi_\gamma (t,{\bf x})$
one can expand square root operators which we meet on the l.h.s. and on
the r.h.s. of this equation in the Taylor series.
Just keeping leading terms we arrive to the non-relativistic equation
%
\begin{equation}
\left( i\partial_t + \frac{1}{2m} \, \nabla^2 \right)
\psi_\gamma (t,{\bf x}) =
- \frac{1}{2m} \, e^{imt} J_\gamma (t,{\bf x})
\, ,
\label{4:5}
\end{equation}
where we skipped all terms of the order $1/c^2$ and higher, they
serve as relativistic corrections.
A general scheme to obtain the non-relativistic equation of motion
to any order of relativistic corrections from
the relativistic equation in the presence of an external
field was elaborated in \cite{anch97}.

\section{ Initial conditions as an external current }
\label{append:initial}
\subsection{ Cauchy problem for Scr\"odinger equation }

In this section we represent the initial conditions of a differential
equation as an external current (the generalized Cauchy problem
\cite{vladimirov}).
We consider a differential equation which contains a first derivative
with respect to time.
To be specific let us take the Schr\"odinger equation
\begin{equation}
\left( i\partial_t - \hat{H}({\bf x}) \right) \psi(x)=0
\label{c:1}
\end{equation}
with initial condition: $\psi(t=0,{\bf x})=\Phi({\bf x})$.
To solve eq.~(\ref{c:1}) we define a new wave function
$\bar{\psi}(t,{\bf x})$ by extension of
$\psi(t,{\bf x})$ for $t < 0$
in the following way
%
\[ \bar{\psi}({\bf x},t) = \left\{ \begin{array}{ll}
                   \psi({\bf x},t),  & \mbox{if $t \ge 0$} \, , \\
                    0,                      & \mbox{if $t < 0$} \, ,
                                    \end{array}
                            \right.  \]
which obviously satisfies eq.~(\ref{c:1}) if $\psi(x)$ is a solution.
As a next step we make the Fourier transformation of eq.~(\ref{c:1})
separating the integral over time in two parts,
$\int^\infty_{-\infty} dt F(t)
=\int^{0^-}_{-\infty} dt F(t)+\int^\infty_0 dt F(t)$.
Then, (\ref{c:1}) reads
\begin{eqnarray}
&&
\int^\infty_{-\infty} dt \, e^{i \omega t} i\partial_t \bar{\psi}(t,{\bf x})
- \hat{H}({\bf x})
  \int^\infty_{-\infty} dt \, e^{i \omega t} \bar{\psi}(t,{\bf x})
\nonumber \\
&&=
i\, e^{i \omega t}\psi(t,{\bf x})\Big|^\infty_0
  - i^2 \omega \int^\infty_0 dt  \, e^{i \omega t} \psi(t,{\bf x})
  - \hat{H}({\bf x}) \bar{\psi}(\omega,{\bf x})
=0
\, ,
\label{c:2}
\end{eqnarray}
where
$\bar{\psi}(\omega,{\bf x})
=\int^\infty_{-\infty} dt \, e^{i \omega t} \bar{\psi}(t,{\bf x}).$
After integration by parts one can insert the function $\Phi({\bf x})$,
which represents the initial condition, to eq.~(\ref{c:2}) keeping in
mind that $e^{i \omega t}\psi(t,{\bf x})\Big|_{t \to \infty} \to 0.$
Then, for the Fourier components we obtain equation
\begin{equation}
\left( \omega - \hat{H}({\bf x}) \right) \bar{\psi}(\omega,{\bf x})
= i \, \Phi({\bf x})
\, .
\label{c:3}
\end{equation}

Let us make the inverse Fourier transformation of eq.~(\ref{c:3}) with
making use of the equality:
$$
\int^\infty_{-\infty} \frac{d\omega}{2\pi} \,
e^{-i \omega t} \omega \, \bar{\psi}(\omega,{\bf x})
=
 i\partial_t \int^\infty_{-\infty} \frac{d\omega}{2\pi} \,
e^{-i \omega t} \, \bar{\psi}(\omega,{\bf x})
=
 i \, \partial_t \, \bar{\psi}(t,{\bf x})
.$$
Then, one obtains equation in time-space representation
\begin{equation}
\left(  i\partial_t - \hat{H}({\bf x}) \right) \bar{\psi}(t,{\bf x})
= i \, \Phi({\bf x}) \, \delta(t)
\, .
\label{c:4}
\end{equation}
The same equation is valid for the function $\psi(t,{\bf x}).$
It is an interesting result because we reduce the initial condition
for differential equation (\ref{c:1}) to impulse current which stands
now on the r.h.s. of equation (\ref{c:4}).

One can define the Green's function
%
$\left(  i\partial_t - \hat{H}({\bf x}) \right) G(x-y)=\delta ^4(x-y)$.
%
With a help of the Green's function one can write solution of
eq.~(\ref{c:4})
\begin{equation}
\psi(t,{\bf x})
=i \int d^3y \, G(t-t_0,{\bf x}-{\bf y}) \, \Phi({\bf y})
\, .
\label{c:6}
\end{equation}
On the other hand, it is solution of the Cauchy problem which was
formulated as eq.~(\ref{c:1}) with initial condition.
Integral on the r.h.s. of (\ref{c:6}) represents propagation of the
initial "excitation" $\Phi({\bf y})$ which exists at time $t=t_0$ to
other spatial points ${\bf x}$ during time interval $(t-t_0)$.

In free case in the non-relativistic limit the Green's function
$G_0(x-y)$ satisfies equation
$\left( i\partial_t +  \nabla^2/2m \right)
G_0(x-y)=\delta ^4(x-y)$,
and it reads
\begin{eqnarray}
 G_0(x)=
 \int \frac{d^4k}{(2\pi )^4}\,
   \frac{\displaystyle e^{-i k\cdot x}}{k^0-\omega ({\bf k})+i\epsilon }
=-i \, \theta (x^0) \left(\frac{m}{2\pi i\, x^0}\right)^{3/2}
  \exp{ \left[ i \frac{m{\bf x}^2}{2 x^0} \right] }
\, ,
\label{c:8}
\end{eqnarray}
where $\omega ({\bf k})={\bf k}^2/2m$.
Note, the non-relativistic Green's functions have the following group
property
\begin{equation}
\int d^3y \, G_0(x-y)\, G_0(y-z)
= -i \,  G_0(x-z)
\, .
\label{c:9a}
\end{equation}
So, using the explicit expression of the free Green's function (\ref{c:8})
one can write solution of the Cauchy problem accumulated in
eq.~(\ref{c:4}) in the following form
\begin{eqnarray}
\psi(t,{\bf x})
&=&  \theta (t) \int \frac{d^3k}{(2\pi )^3}\, d^3y
  e^{-i \, \omega({\bf k})t + i \, {\bf k}\cdot ({\bf x}-{\bf y})}
  \, \Phi({\bf y})
= \theta (t) \int \frac{d^3k}{(2\pi )^3}\,
  e^{-i \, \omega({\bf k})t + i \, {\bf k}\cdot {\bf x} }
  \, \Phi({\bf k})
\nonumber \\
&=&  \theta (t) \left(\frac{m}{2\pi i t}\right)^{3/2}
  \int d^3y \, \exp{ \left[ i \frac{m({\bf x}-{\bf y})^2}{2 t} \right] }
   \, \Phi({\bf y})
\, ,
\label{c:9}
\end{eqnarray}
where
$\Phi({\bf k})= \int d^3y  \exp{( -i \, {\bf k}\cdot {\bf y} ) }
  \, \Phi({\bf y}).$

We turn now to another way of solution of the Cauchy problem
(\ref{c:1}).
Formally one can write the solution of the problem as
\begin{equation}
\psi(t,{\bf x})
=
\theta(t-t_0) \, e^{-i \hat{H}(\hat{{\bf p}},\,{\bf x})(t-t_0) }
 \Phi({\bf x})
\, ,
\label{c:10}
\end{equation}
where $\hat{{\bf p}}= -i\hbar \partial/\partial{\bf x}$ is the momentum
operator  and
$\psi(t=t_0,{\bf x})=\Phi({\bf y})$, in what follows we adopt $t_0=0.$
We are going to give the formal solution (\ref{c:10})
in coordinate representation and then compare it with (\ref{c:9}).
Indeed, in coordinate representation (\ref{c:10}) reads
\begin{equation}
\psi(t,{\bf x})
=
\theta(t) \,
\int \frac{d^3p}{(2\pi)^3} \, \frac{d^3p'}{(2\pi)^3} \, d^3x' \,
  \langle {\bf x}|{\bf p}\rangle
  \langle {\bf p}|e^{-i \hat{H}(\hat{{\bf p}},{\bf x})t }
    |{\bf p}' \rangle \,
  \langle {\bf p}'|{\bf x}' \rangle \, \Phi({\bf x}')
\, ,
\label{c:10a}
\end{equation}
where $\langle {\bf x}|{\bf p} \rangle=\exp{(i {\bf p}\cdot {\bf x})}$
is the eigen function of the momentum operator.
In free case the matrix elements of the Hamiltonian,
$\hat{H}_0(\hat{{\bf p}},{\bf x})= \hat{{\bf p}}^2/2m$,
looks like,
$\langle {\bf p}|e^{-i \hat{H}_0(\hat{{\bf p}},{\bf x})t }
    |{\bf p}' \rangle \,
=
(2\pi)^3 \, \delta^3({\bf p}- {\bf p}') \, e^{-i \omega({\bf p})t }$,
where $\omega({\bf p})={\bf p}^2/2m.$
Finally, for free case we can write (\ref{c:10}) in coordinate
representation
\begin{equation}
\psi(t,{\bf x})
=
\theta(t) \,
\int \frac{d^3p}{(2\pi)^3} \, d^3x' \,
e^{-i \omega({\bf p})t + i{\bf p}\cdot ({\bf x}- {\bf x}') }
\Phi({\bf x}')
\, .
\label{c:18}
\end{equation}
We see that this expression coincides with (\ref{c:9}).
So, we find that it does not matter in what approach one solves the
Cauchy problem, with the help of the Green's function or using
evolution operator, both approaches give the same result.

\subsection{ Cauchy problem for relativistic equation }

We consider the case of free scalar field which can be described
by the Klein-Gordon equation
\begin{equation}
(\partial_\mu \partial^\mu + m^2)\Psi(x)=0
\, ,
\label{c:19}
\end{equation}
which is supplemented by the initial conditions
\begin{equation}
\Psi(t=0,{\bf x})=\Phi_0({\bf x})\, ,
\ \ \ \ {\rm and} \ \ \ \
\frac{\partial \Psi(t=0,{\bf x})}{\partial t}= \Phi_1({\bf x})
\, .
\label{c:20}
\end{equation}
We are going to show how these initial conditions can be inserted into
eq.~(\ref{c:19}) as a specific current.
As the first step let us extend the function $\Psi(t,{\bf x})$
(we define the function equals to zero for negative times)
\[ \bar{\Psi}({\bf x},t) = \left\{ \begin{array}{ll}
                   \Psi({\bf x},t),  & \mbox{if $t \ge 0$} \, , \\
                    0,                      & \mbox{if $t < 0$} \, .
                                    \end{array}
                            \right.  \]
We make the Fourier transformation of eq.~(\ref{c:19}) separating the
integral over time in two parts, $\int^\infty_{-\infty} dt F(t)
=\int^{0^-}_{-\infty} dt F(t)+\int^\infty_0 dt F(t)$.
Then, one obtains
\begin{eqnarray}
\int^\infty_{-\infty} dt \, e^{i \omega t} (\partial_\mu \partial^\mu + m^2)\Psi(x)
&=&
\int^\infty_{-\infty} dt \, e^{i \omega t} \partial_t^2 \bar{\Psi}(t,{\bf x})
+ \left(- \partial_{\bf x}^2 + m^2 \right)
  \int^\infty_{-\infty} dt \, e^{i \omega t} \bar{\Psi}(t,{\bf x})
\nonumber \\
&& \hspace{-1cm}
=  e^{i \omega t} \frac{\partial \Psi(t,{\bf x})}{\partial t} \Big|^\infty_0
  - i \omega \int^\infty_0 dt  \, e^{i \omega t}
      \frac{\partial \Psi(t,{\bf x})}{\partial t}
  + \left(- \partial_{\bf x}^2 + m^2 \right) \bar{\Psi}(\omega,{\bf x})
\nonumber \\
&& \hspace{-1cm}
= - \Phi_1({\bf x}) + i \omega \Phi_0({\bf x})
  - \omega^2 \bar{\Psi}(\omega,{\bf x})
  + \left(- \partial_{\bf x}^2 + m^2 \right) \bar{\Psi}(\omega,{\bf x})
,
\label{c:21}
\end{eqnarray}
where
$\bar{\Psi}(\omega,{\bf x}) =\int^\infty_{-\infty} dt \, e^{i
\omega t} \bar{\Psi}(t,{\bf x}).$
After two integrations by parts we insert the functions
$\Phi_0({\bf x})$ and $\Phi_1({\bf x})$, which represent the initial
conditions, to eq.~(\ref{c:21}) keeping in mind that
$e^{i \omega t}\Psi(t,{\bf x})\Big|_{t \to \infty} \to 0.$
Then, for the Fourier components we obtain the following equation
\begin{equation}
\left(- \omega^2 - \partial_{\bf x}^2 + m^2 \right) \bar{\Psi}(\omega,{\bf x})
=  \Phi_1({\bf x}) - i \omega \Phi_0({\bf x})
\, .
\label{c:22}
\end{equation}

We make the inverse Fourier transformation of eq.~(\ref{c:22}) with
making use of the equalities:
$$
\int^\infty_{-\infty} \frac{d\omega}{2\pi} \,
e^{-i \omega t} (-\omega^2) \, \bar{\Psi}(\omega,{\bf x})
= \partial_t^2  \bar{\Psi}(t,{\bf x})\, , \ \ \ \
\int^\infty_{-\infty} \frac{d\omega}{2\pi} \,
 (-i \, \omega) \, e^{-i \omega t}
=  \frac{d}{dt} \, \delta(t) \, .
$$
Then, one obtains equation in the space-time representation
\begin{equation}
\left(  \partial_t^2 - \partial_{\bf x}^2 + m^2 \right) \bar{\Psi}(t,{\bf x})
=  \Phi_1({\bf x}) \, \delta(t) + \Phi_0({\bf x}) \,  \frac{d}{dt} \, \delta(t)
\, .
\label{c:23}
\end{equation}
The same equation is valid for the function $\Psi(t,{\bf x})$ for times, $t\ge 0$,
\begin{equation}
(\partial_\mu \partial^\mu + m^2)\Psi(x)
=  \Phi_1({\bf x}) \, \delta(t) + \Phi_0({\bf x}) \, \delta'(t)
\, .
\label{c:24}
\end{equation}
So, we represent the initial conditions for the
differential equation (\ref{c:19}) as a current which stands
on the r.h.s. of equation (\ref{c:24}).
The use of the function $\delta'(t)$ is common:
$\int dt F(t)\delta'(t)=-\int dt \frac{dF(t)}{dt}\delta(t).$

To solve eq.(\ref{c:24}) we use the Green's functions $G_R(x-y)$ (or
$G_F(x-y)$).
Because in further consideration we look for solutions
$\Psi(x^0,{\bf x})$ at asymptotic times, $x^0 \to \infty$, just a
piece of the propagator $G_R(x-y)$ which carries the positive defined
frequencies really gives contribution.
(Because of that it does not matter what kind of the Green's function
should be used, retarded or causal one.)
So, with the help of the propagator $G_R(x-y)$ which is
defined as, $(\partial_\mu \partial^\mu + m^2)G_R(x-y)=\delta ^4(x-y)$,
we can write solution of eq.~(\ref{c:24})
\begin{eqnarray}
\Psi(x)
&=&
\int  d^4y \, G_R(x-y)
  \left[  \Phi_1({\bf y}) \, \delta\left(y^0 \right)
  + \Phi_0({\bf y}) \, \delta'\left(y^0\right)
  \right]
\nonumber \\
&=&
\int  d^4y \, \delta\left(y^0\right)
\left[  G_R(x-y) \, \frac{\partial \Psi(y^0,{\bf y})}{\partial y^0}
      - \frac{\partial G_R(x-y) }{\partial y^0}\, \Psi(y^0,{\bf y})
  \right]
 \, .
\label{c:26}
\end{eqnarray}
The integration in (\ref{c:26}) is going on the space-like
hypersurface $y^0=0$.
For arbitrary  hypersurface $\sigma$ it can be written in the
covariant form in the following way (see, for instance,
\cite{schweber}, ch.~7)
\begin{eqnarray}
\Psi(x)
=
\int_{(\sigma)}  d\sigma^\mu(y) \, G_R(x-y) \,
\frac{\stackrel{\leftrightarrow}{\partial } }{\partial y^\mu} \,
\Psi(y)
\label{c:27}
\end{eqnarray}
with $\sigma^\mu(y)$ as the space-like hypersurface on which the
initial conditions $\Psi(y)$ and
$\partial \Psi(y)/\partial y^\mu $ are given, i.e. these functions
under the integral are defined when $y \in \sigma$.
By definition,
$f_1(t) \stackrel{\leftrightarrow}{\partial }_t f_2(t) \equiv
f_1(t) \partial_t f_2(t)-\partial_t(f_1(t))  f_2(t).$


We derive now one more relation which is useful in the consideration.
One can write the Green's functions in the following way
\begin{eqnarray}
 G_R(x-y)
=\theta \left( x^0-y^0 \right) \left[ G^{(+)}(x-y)- G^{(-)}(x-y) \right]
 \, .
\label{c:29}
\end{eqnarray}
where
\begin{eqnarray}
G^{(\pm)}(x-y)
= i \int \frac{d^3k}{(2\pi )^3 2\omega ({\bf k})} \,
  f^{(\pm)}_k(x) \,f^{(\pm),*}_k(y)
\label{c:30}
\end{eqnarray}
with $f^{(\pm)}_k(x)=e^{\mp k\cdot x}$, which obey the orthogonal
relations.
It is obvious that the functions $G^{(+)}(x-y)$ and $G^{(-)}(x-y)$
satisfy the Klein-Gordon equation,
$(\partial_\mu \partial^\mu + m^2)G^{(\pm)}(x-y)=0 \, .$
With taking into account normalization,
$\int d^3x \,
f^{(+),*}_{k}(x) \, i \stackrel{\leftrightarrow}{\partial }_{x^0}
  f^{(+)}_{p}(x)
=
(2\pi )^3\, 2\omega ({\bf k})\, \delta ^3({\bf k}-{\bf p}),$
one obtains
\begin{equation}
\int  d^4z \, \delta\left(z^0\right)
\left[  G^{(+)}(x-z) \, \frac{\partial \, G^{(+)}(z-y)}{\partial z^0}
      - \frac{\partial G^{(+)}(x-z) }{\partial z^0 }\, G^{(+)}(z-y)
  \right]
= G^{(+)}(x-y)
 \, .
\label{c:31}
\end{equation}
This can be written for arbitrary  space-like hypersurface $\sigma$
in the covariant form \cite{schweber}
\begin{eqnarray}
G^{(+)}(x-y)
=
\int_{(\sigma)}  d\sigma^\mu(z) \,G^{(+)}(x-z) \,
\frac{\stackrel{\leftrightarrow}{\partial } }{\partial z^\mu} \,
G^{(+)}(z-y)
 \, .
\label{c:32}
\end{eqnarray}
%


\end{document}